\documentstyle[11pt,paspconf,epsf]{article}

\begin{document}

\title{The CMBR and the Seeds of Galaxies}
\author{Edward L. Wright}
\affil{UCLA Astronomy, PO Box 951562, Los Angeles CA 90095-1562}

\begin{abstract}
The Cosmic Microwave Background Radiation (CMBR) is the radiation left over
from the hot Big Bang.  Its blackbody spectrum and small anisotropy provide
clues about the origin and early evolution of the Universe.
In particular, the spectrum of the CMBR rules out many non-gravitational models
of structure formation, and the anisotropy of the CMBR provides a measure of
the gravitational potential at the time of last scattering, 
about $10^{5.5}$~years after the Big Bang.  
The density inhomogeneities needed to produce
the gravitational potential perturbations traced by the CMBR have grown to
become the galaxies, clusters of galaxies, and superclusters that we see today.
\end{abstract}


\section{Introduction}

The observations made by the {\sl COBE}
\footnote{
The National Aeronautics and Space Administration/Goddard Space Flight
Center (NASA/GSFC) is responsible for the design, development, and
operation of the Cosmic Background Explorer (COBE).
Scientific guidance is provided by the COBE Science Working Group.
GSFC is also responsible for the development of the analysis software and
for the production of the mission data sets.
}
project (Boggess {\it et al.\/} 1992)
were part of a long history of discoveries about the Universe.
The replacement of Ptolemy's geocentric cosmology with the heliocentric
cosmology of Copernicus was a first step in moving humanity from a unique and
special position in the cosmos to a typical location.  The replacement
of Kapetyn's galaxy with Shapley's larger system put the Solar System
quite far offcenter in the Universe.  
But studies of the extragalactic nebulae showed
that the Universe was far larger than any one galaxy, and that the
position of our Milky Way in the Universe was not unique.
Hubble (1929) found a linear relationship between the distance to a galaxy
and its recession velocity measured by its redshift.  This observation
fit in with expanding models for the Universe that had been worked out using
the theory of general relativity, including the exponentially expanding
but zero density de Sitter model, the 
critical density Einstein--de~Sitter model, and the
more general models found by Friedmann.
The creation of the elements in Friedmann models was studied by
Gamow (1946) and led to the realization that the Universe had to be hot
during its early phases, and that this radiation should still be present
with a current temperature of a few degrees Kelvin (Alpher \&
Herman (1948) compute 5 K.)
But the lack of stable nuclei with A = 5 or A = 8 ultimately meant that
the Gamow model for the creation of all the elements could in fact
only produce hydrogen, helium and a trace of lithium (Copi, Schramm \& Turner
1995).
The de Sitter model was further developed by Hoyle and by Bondi \& Gold
into the Steady State model, in which a continuous creation of matter
allowed a finite (and constant) density even though the Universe expands
exponentially.
And Burbidge, Burbidge, Fowler \& Hoyle (1957) found that the elements 
heavier than
helium could be made from hydrogen in stars, but later work showed
that the helium to heavy
element ratio produced by stars was less than the observed ratio.
This ``helium'' problem led Dicke, Peebles, Roll \& Wilkinson (1965)
to search for the CMBR,
but they were ``scooped'' by Penzias \& Wilson (1965) who had seen the
radiation because it was a significant contributor to the total noise in their
ultra-sensitive receiver at Bell Labs.
Previously Dicke, Beringer, Kyhl \& Vane (1946) had placed an upper limit of 
$< 20$~K on the CMBR radiation, so it would have been easy to detect the CMBR
at any time after 1945 with the microwave technology developed during WW~II.
Even earlier Adams (1941) calls attention to observations of interstellar CN 
molecules in a rotationally excited state (McKellar 1941), 
and this is the first observation 
of the CMBR.  But it was only after the discovery of the CMBR by Penzias \&
Wilson (1965) that the rotational excitation of interstellar CN was developed
into the most accurate groundbased measurement of the temperature of the CMBR,
giving $T_\circ = 2729^{+23}_{-31}$~mK (Roth, Meyer \& Hawkins 1993).

The existence of the CMBR rules out the Steady State model of the Universe,
for the Universe today is not the isothermal and opaque Universe necessary
to produce a blackbody spectrum.
Fortunately one of the main motivations behind the Steady State -- the
discrepancy between the expansion age of the Universe $1/H_\circ$
and the measured ages of the oldest things in the Universe --
has largely disappeared due to the new {\sl HIPPARCOS\/}
subdwarf parallaxes.  These put the globular clusters further away, and
hence the stars at the main sequence turnoff are more luminous and thus
younger.  Reid (1997) obtains an age of $t_\circ = 12\pm1$~Gyr
for the oldest globular clusters.
The {\sl HIPPARCOS\/} recalibration of the Cepheid PL relation will probably
lower the Riess, Press \& Kirshner (1966) value of 
$H_\circ = 64 \pm 6$~km/sec/Mpc by a few percent, giving the dimensionless
product of $H_\circ$ and $t_\circ$ 
a value of $H_\circ t_\circ = 0.76\pm0.1$ which is compatible with the
$2/3$ predicted by the $\Omega = 1$ Einstein-de~Sitter model.

The expansion of the Universe will not change the blackbody character
of the CMBR.  While the redshift reduces the frequency of the photons 
(and hence the color temperature) by 
a factor $a$, the expansion of the Universe reduces the number 
density of photons by by a factor of $a^3$, so the energy density
is proportional to the color temperature to the fourth power, and a
blackbody continues to look like a blackbody.
Thus in the evolving Universe of the Big Bang model, if the isothermal
and opaque conditions conditions necessary to produce a blackbody
exist at early times, and if there is no substantial transfer of energy
into the CMBR at later times, then the spectrum of the CMBR will be
very close to a blackbody, and the magnitude of the deviations can be
used to determine the nature of any energy transfers into the CMBR.

The existence of the CMBR also drastically changed the theory of structure
formation in the Universe.  
During the first few hundred thousand years after the Big Bang, photons
and baryonic matter were strongly coupled by Thompson scattering.
At a time $10^{5.4}$ years\footnote{
With $H_\circ = 50,\;\Omega = 1$.}
after the Big Bang, at $z \approx 1360$, the
temperature has fallen to the point where helium and then hydrogen 
have 50\% (re)combined into transparent gases$^5$.
The surface of last scattering ($\partial \tau/\partial \ln(1+z) = 1$)
occurs later, at $z_r \approx 1160$ or $10^{5.7}$ years after the
Big Bang.
The electron scattering which had impeded the free
motion of the CMBR photons until this epoch is removed, and the photons stream
across the Universe.  Before recombination, the radiation field at any point
was constrained to be nearly isotropic because the rapid scattering
scrambled the directions of photons.  
The radiation field was not required to be homogeneous, 
because the photons remained approximately fixed in comoving
co-ordinates.  
After recombination, the free streaming of the photons has the
effect of averaging the intensity of the microwave background over a region
with a size equal to the horizon size.  Thus after recombination any
inhomogeneity in the microwave background spectrum is smoothed out.  
Note that this
inhomogeneity is not lost: instead, it is converted into anisotropy.  
When we
study the isotropy of the microwave background, we are looking back to the
surface of last scattering $\approx 10^{5.5}$ years after the Big Bang.  
But the hot spots
and cold spots we are studying existed as inhomogeneities in the Universe
before recombination.  
Since the $7^\circ$ beam used by the DMR instrument on
{\sl COBE\/} is larger than the horizon size at recombination, 
these inhomogeneities cannot be constructed in a causal fashion during 
the epoch before recombination in the standard Big Bang model.   
Instead, they must be installed ``just so'' in the initial conditions.  
In the inflationary scenario (Starobinsky 1980, Guth 1981)
these large scale structures were once smaller than the horizon size during 
the inflationary epoch, but grew to be much larger than the horizon.  Causal
physics acting $10^{-35}$~seconds after the Big Bang can produce the
large-scale inhomogeneities studied by the DMR.

A natural consequence of the inflationary scenario is the production
of a perturbation spectrum which approximates the Harrison-Zel'dovich
spectrum (Harrison 1970; Zel'dovich 1972; Peebles \& Yu 1970).  When
expressed as the power spectrum of the density contrast, this give
\begin{equation}
P(k) = k^n\quad\quad\mbox{with}\quad\quad n = 1
\end{equation}
The density contrast on a length scale $\lambda = 2\pi/k$ is
given by $\delta = \sqrt{k^3 P(k)}$ and is thus proportional to
$\lambda^{-2}$ in this model.  As a result, the gravitational potential
fluctuations are independent of the length scale: 
$\Delta\phi \propto \lambda^0$.  This is known as ``equal power on all
scales''. 

Because the density contrast of perturbations only
grows in proportion to the size of the Universe, the existence of 100\% density contrasts now implies the presence of 0.1\% density contrasts at the epoch of
recombination.  Since the density of photons is proportional to $T^3$,
this leads to a prediction of $\Delta T/T = 3 \times 10^{-4}$
(Silk 1968).
The length scale which now has $\Delta\rho/\rho = 1$ is approximately
$8h^{-1}$~Mpc where $h = H_\circ/(100$~km/sec/Mpc).  This translates
into an angle of $\theta = (800\Omega\;\mbox{km/sec})/(2c) = 5^\prime$.
The predicted 1 mK temperature fluctuations on these scales were looked
for and not seen.

However, observations of clusters of galaxies suggest that most of the
matter in the Universe is non-luminous.  If this {\em dark} matter
does not emit, scatter or absorb light, then it will be free to collapse
into gravitational potential wells while the ordinary, or {\em baryonic},
matter is prevented from collapsing by the radiation pressure of the 
CMBR.  Thus the dark matter can have a density contrast of 0.1\%
as recombination, while the ordinary matter and photons have a much
smaller density contrast.  Peebles (1982) calculated the expected
$\Delta T$ in this model, now called Cold Dark matter (CDM), and found
tens of $\mu$K instead of 1 mK.

\begin{figure}
\plotone{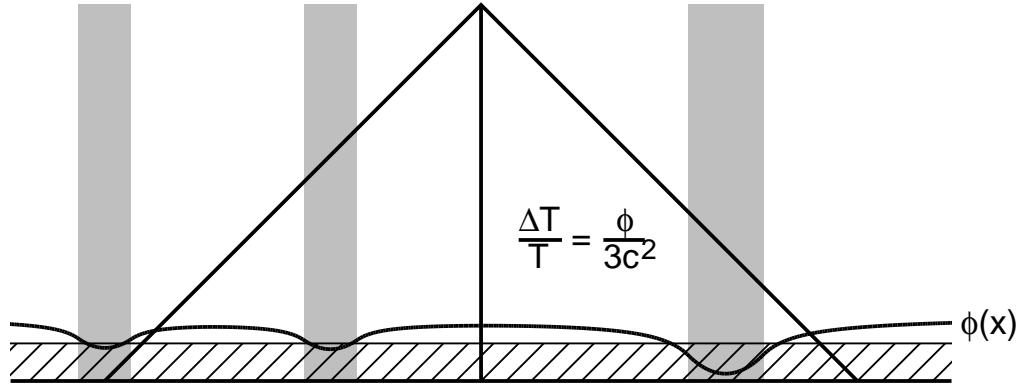}
\caption{The Sachs-Wolfe effect produces cool spots in regions
where a dense lump of matter produces a negative potential.
This is a conformal space-time diagram,\label{fig:saxwolfe}}
\end{figure}

\begin{figure}
\plotone{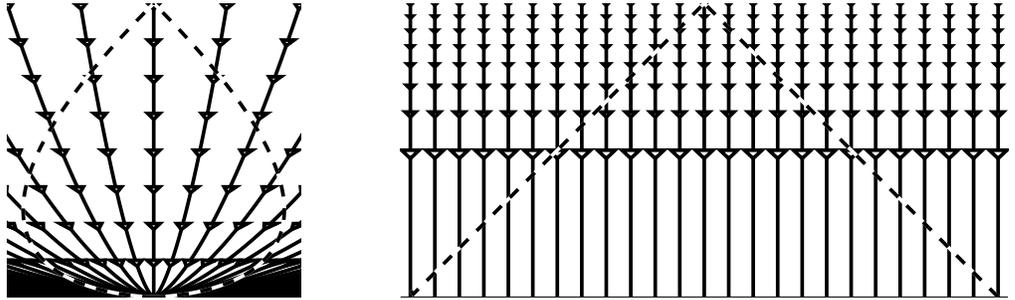}
\caption{Comparison of an ordinary space-time diagram for an $\Omega = 1$
model on the left to a conformal space-time diagram in the right.
The dashed curves are the past light cone of the central observer.
\label{fig:critconf}}
\end{figure}

\begin{figure}
\plotone{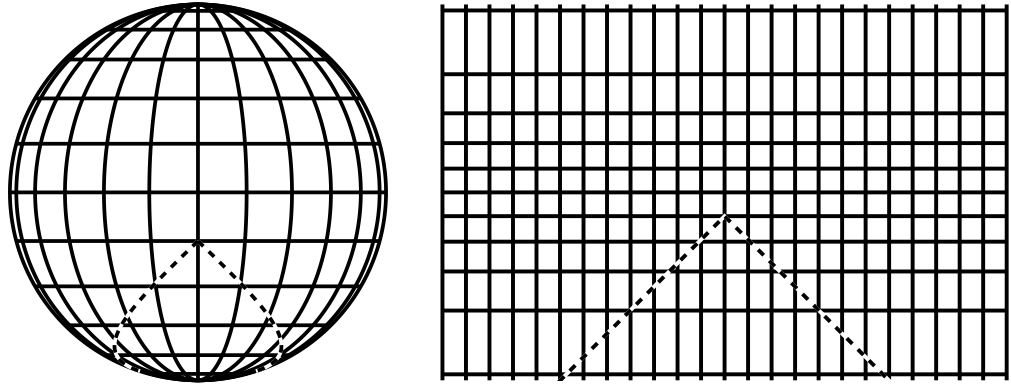}
\caption{An analogy to Figure \protect\ref{fig:critconf}:
a side view of a sphere on the left, and a conformal Mercator map
on the right.  The dashed curves are constant SE and SW courses.
\label{fig:combosm2}}
\end{figure}

Another way that matter can affect the CMBR was found by Sachs \&
Wolfe (1967).  They found that a gravitational potential perturbation
produces a temperature fluctuation of $\Delta T/T = \Delta\phi/(3c^2)$.
Since the potential is an integral over the density, this effect
dominates at larger angular scales.  Figures \ref{fig:saxwolfe},
\ref{fig:critconf} and \ref{fig:combosm2} illustrate the Sachs-Wolfe
effect.

\section{Spectrum}

\subsection{FIRAS Observations}

The Far InfraRed Absolute Spectrophotometer instrument on {\sl COBE\/} is
a polarizing Michelson (Martin \& Puplett 1970) interferometer.  
The optical layout
is symmetrical, and it has two inputs and two outputs.  
If the two inputs
are denoted SKY and ICAL, then the two outputs, which are denoted LEFT and
RIGHT, are given symbolically as 
\begin{eqnarray}
{\rm LEFT} & = & {\rm SKY} - {\rm ICAL}\\
{\rm RIGHT} & = & {\rm ICAL} - {\rm SKY}
\end{eqnarray}
The FIRAS has achieved its incredible sensitivity to small deviations from a
blackbody spectrum by connecting the ICAL input to an internal calibrator, a
reference blackbody that can be set to a temperature close to the
temperature $T_\circ$ of the sky.  Thus this ``absolute'' spectrophotometer
is so successful because it is differential.
In addition, each output is further divided by a dichroic beamsplitter into a
low frequency channel (2-21 cm$^{-1}$) and a high frequency 
channel (23-95 cm$^{-1}$).
Thus there are four overall outputs.  These are labeled LL (left low)
through RH (right high).

Since the FIRAS is a Michelson interferometer, the spectral data are obtained
in the form of interferograms.  Thus the LEFT output is approximately
\begin{equation}
I_L(x) = \int_0^\infty \cos(2\pi\nu x) 
G(\nu) \left(I_\nu + \sum_i \epsilon_i(\nu) B_\nu(T_i) + U_\nu \right) d\nu
\end{equation}
The index $i$ above runs over all the components in the FIRAS that had
thermometers to measure $T_i$.  These include the ICAL, the reference horn that
connects to the ICAL, the sky horn, the bolometer housing, the optical
structure of the FIRAS, and the dihedral mirrors that move to provide the
variation in path length difference $x$.
The $\epsilon_i$'s are the effective emissivities of the various components.
The ICAL itself has $\epsilon \approx -1$, while the sky and reference horns
have $\epsilon$'s of $\pm$ a few percent in the low frequency channels.
The offset term $U_\nu$ was observed during flight to be approximately
$10^{-5}\exp(-t/\tau)B_\nu(T_U)$ with a time constant $\tau$ of two months
and a temperature $T_U \approx 15$~K.
The SKY input $I_\nu$ above can be either the sky or an external calibrator.
The XCAL is a movable re-entrant absorber that can be inserted at the top of
the sky horn.  The combination of sky horn plus XCAL forms a cavity with an
absorptivity known to be $> 0.9999$ from measurements,
and believed to be $> 0.99999$ from calculations.
With the XCAL inserted during periodic calibration runs, the SKY input is
known to be $B_\nu(T_X)$.  By varying $T_X$ and the other $T_i$'s, the
calibration coefficients have been determined (Fixsen {\it et al.} 1994).  

Once the calibration coefficients are known, the sky data can be analyzed to
determine $I_\nu(l,b)$ the intensity of the sky as a function of frequency,
galactic longitude and galactic latitude.  This is a ``data cube''.  The data
from each direction on the sky can be written as a combination of cosmic plus
galactic signals:
\begin{eqnarray}
I_\nu(l,b) & = & e^{-\tau_\nu(l,b,\infty)} 
\left(B_\nu\left(T_\circ + \Delta T(l,b)\right) + \Delta I_\nu \right)\nonumber 
\\
& + & \int e^{-\tau_\nu(l,b,s)} j_\nu(l,b,s) ds
\end{eqnarray}
where $\tau_\nu(l,b,s)$ is the optical depth between the Solar system and 
the point at distance $s$ in the direction $(l,b)$ at frequency $\nu$,
$\Delta I_\nu$ is an isotropic cosmic distortion, and
$\Delta T(l,b)$ is the variation of the background temperature around its mean
value $T_\circ$.  This equation can be simplified because the optical
depth of the galactic dust emission is always small in the
millimeter and sub-millimeter bands covered by FIRAS.  
\begin{equation}
I_\nu(l,b)  \approx B_\nu\left(T_\circ + \Delta T(l,b)\right) + \Delta I_\nu
 +  \int j_\nu(l,b,s) ds
\end{equation}
Even in the optically
thin limit, some restrictive assumptions about the galactic emissivity
$j_\nu$ are needed, since the galactic intensity
$\int j_\nu(l,b,s) ds$ is a function of three variables, just like the
observed data.  The simplest reasonable model (Wright {\it et al.} 1991) 
for the galactic emission is
\begin{equation}
\int j_\nu(l,b,s) ds = G(l,b) g(\nu). 
\end{equation}
This model assumes that the shape of the galactic spectrum is independent
of direction on the sky.  It is reasonably successful except that the 
galactic center region is clearly hotter than the rest of the galaxy.
The application of this model proceeds in two steps.  The first step
assumes that the cosmic distortions vanish, and that an approximation
$g_\circ(\nu)$ to the galactic spectrum is known.  A least squares fit
over the spectrum in each pixel then gives the maps $\Delta T(l,b)$
and $G(l,b)$.  The high frequency channel of FIRAS is used to derive
$G(l,b)$ because the galactic emission is strongest there.  An alternative
way to derive $G(l,b)$ is to smooth the DIRBE map at 240 $\mu$m to the FIRAS
$7^\circ$ beam.  A modification of this DIRBE method has been used in the 
latest FIRAS spectral
results (Fixsen {\it et al.} 1996) where it is assumed that
\begin{equation}
\int j_\nu(l,b,s) ds = G_1(l,b) g_1(\nu) + G_2(l,b) g_2(\nu)
\end{equation}
and the maps $G_1$ and $G_2$ are derived from the DIRBE Band 9 (140 $\mu$m)
and Band 10 (240 $\mu$m) maps.

\begin{figure}
\plotone{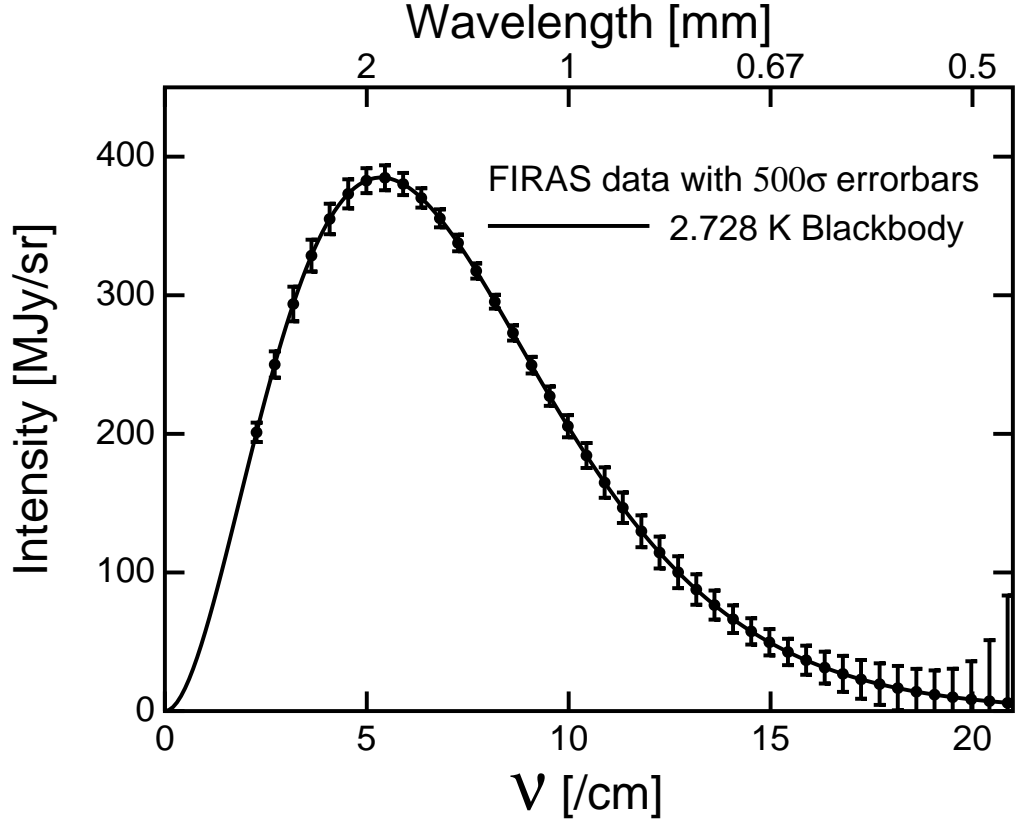}
\caption{Spectrum of the CMBR compared to a blackbody.  The error
bars have been multiplied by 500 to make them visible.
\label{fig:spectrum}}
\end{figure}

The second step in the galactic fitting then derives
spectra associated with the main components of the millimeter wave sky:
the isotropic cosmic background, the dipole anisotropy, and the galactic
emission(s).  This fit is done by fitting all the pixels (except for 
the galactic center region with $|b| < 20$ and $|l| < 40$)
at each frequency to the form
\begin{equation}
I_\nu(l,b) = I_\circ(\nu) + D(\nu)\cos\theta + G_1(l,b) g_1(\nu) +
G_2(l,b) g_2(\nu).
\end{equation}
The final spectrum reported for deviations of the CMBR from a blackbody
is given by 
\begin{equation}
\Delta I_\nu = I_\circ(\nu) - B_\nu(T_\circ) - G_1 g_1(\nu) - G_2 g_2(\nu)
\end{equation}
where the parameters $T_\circ$, $G_1$ and $G_2$ are adjusted to minimize
the $\chi^2$ of the fit.  The noise varies with frequency and noise
in adjacent points is somewhat anti-correlated because of the offcenter
scan used when taking the interferograms.
Figure \ref{fig:spectrum} shows the isotropic spectrum $I_\circ(\nu)$
compared to a 2.728~K blackbody.

The absolute temperature of the cosmic background, $T_\circ$, can be
determined two ways using FIRAS.  The first way is to use the readings
of the germanium resistance thermometers in the XCAL when the XCAL
temperature is set to match the sky.  This gives $T_\circ = 2.730$~K.
The second way is to measure the frequency of the peak of 
$\partial B_\nu/\partial T$ by varying $T_X$ a small amount
around the temperature which matches the sky, and then apply
the Wien displacement law to convert this frequency into a
temperature.  This calculation is done automatically by the
calibration software, and it gives a value of 2.726~K for
$T_\circ$.  
Three additional determinations of $T_\circ$ depend on the dipole
anisotropy.
For either FIRAS or DMR, the spectrum of the dipole anisotropy
can be fit to the form
\begin{equation}
D(\nu) = \frac{T_\circ v}{c} \frac {\partial B_\nu(T_\circ)} {\partial T}
\end{equation}
Since the velocity of the solar system with respect to the CMB is not known
{\it a priori}, only the shape and not the amplitude of the dipole spectrum
can be used to determine $T_\circ$.
For FIRAS, this analysis gives $T_\circ = 2.717 \pm 0.007$~K
(Fixsen {\it et al.} 1996),
while for DMR it gives $T_\circ = 2.76 \pm 0.18$~K
(Kogut {\it et al.} 1993).
The DMR data analysis keeps track of the changes in the dipole caused
by the variation of the Earth's velocity around the Sun during the year.
In this case the velocity $v$ is known, so $T_\circ$ can be determined
from the amplitude of the change in the dipole, giving 
$T_\circ = 2.725 \pm 0.02$~K (Kogut {\it et al.} 1996b).
The final adopted value (Fixsen {\it et al.} 1996) is $2.728 \pm 0.004$~K
(95\% confidence),
which just splits the difference between the two methods based on the
FIRAS spectra.
The dipole-based determinations of $T_\circ$ are less precise but provide
a useful confirmation of the spectral data.

\subsection{Interpretation}

For redshifts greater than $z_y = 10^{5.1}/\sqrt{70\Omega_B h^2}$,
the rate of photon frequency diffusion due to Compton scattering
is large:
\begin{equation}
(1+z)\frac{\partial y}{\partial z} = 
\sigma_T n_{e,\circ} \frac {k T_\circ}{m_e c^2} \frac {c}{H}(1+z)^4 > 1
\end{equation}
where the Kompaneets $y$ is defined by
$dy = (k T_e/m_e c^2) n_e \sigma_T c dt$.
A Bose-Einstein distribution with dimensionless chemical potential $\mu$,
$n = 1/(\exp[x+\mu]-1)$, is a fixed point of the Kompaneets (1957)
equation:
\begin{equation}
\frac {\partial n}{\partial y} = x^{-2} \frac{\partial}{\partial x}
\left[x^4\left(n + n^2 + \frac{\partial n}{\partial x}\right)\right]
\end{equation}
where $n$ is the number of photons per mode ($n = 1/(e^x-1)$ for a blackbody)
and $x = h\nu/kT_e$.
Therefore, any distortion created before $z_y$ will be converted into
a $\mu$ distortion by electron scattering.  When expressed as a frequency
dependent brightness temperature with a conserved photon number, the form
of a $\mu$ distortion is
\begin{equation}
T_\nu = T_\circ \left(1 + \mu 
\left[\frac{\zeta(2)}{3\zeta(3)} - x^{-1}\right] + \ldots \right).
\end{equation}
This spectrum has excess energy relative to a blackbody in the amount of
\begin{equation}
\frac {\Delta U}{U} = 
\left(\frac {4\zeta(2)}{3\zeta(3)}-\frac{\zeta(3)}{\zeta(4)}\right) \mu
= 0.714\mu.
\end{equation}

\begin{figure}[t]
\plotone{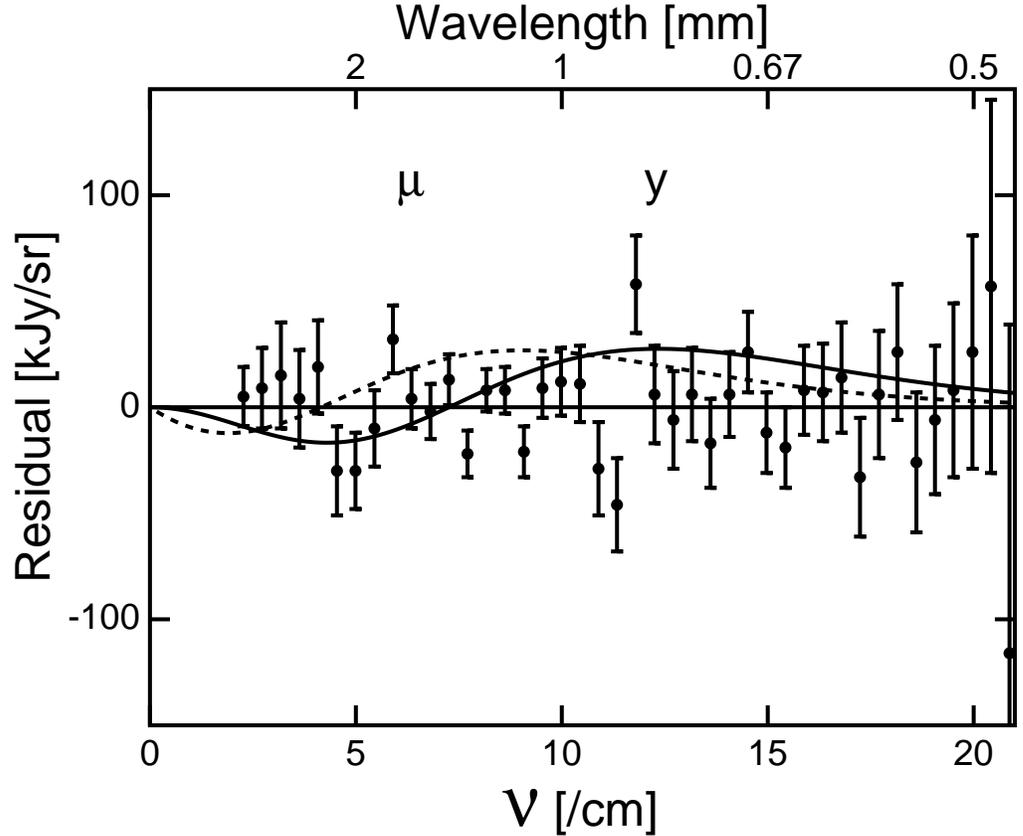}
\caption{The residual after subtracting a blackbody and a constant times
the galactic spectrum from the FIRAS CMBR spectrum.  The 95\% confidence
limits on $\mu$ and $y$ distortions are shown by the dashed and solid
curves.\label{fig:residual}}
\end{figure}

At still higher redshifts, the double photon Compton scattering process
$\gamma + e^- \leftrightarrow \gamma + \gamma + e^-$
can create photons.
A proper consideration (Burigana {\it et al.} 1991) of the interaction
of this photon creation process with the Kompaneets equation
shows that the redshift from which
$1/e$ of an initial distortion can survive is
\begin{equation}
z_{th} = \frac {4.24 \times 10^5}
{\left[\Omega_B h^2 \right]^{0.4}}
\end{equation}
which is $z_{th} = 2.3 \times 10^6$ for the BBNS value of $\Omega_B h^2$.

Other distortions formed at redshifts lower than $z_y$ can survive to the
present.  The distortion formed when a blackbody spectrum is scattered
by hotter electrons is given by a frequency-dependent
temperature (Zel'dovich \& Sunyaev 1969)
\begin{equation}
T_\nu = T_\circ \left[1 + y_D
\left(\frac {x (e^x+1)}{e^x-1} - 4 \right) + \ldots \right].
\end{equation}
where the ``distorting'' $y$ is
\begin{equation}
dy_D = \frac{k (T_e - T_\gamma)}{m_e c^2} n_e \sigma_T c dt.
\end{equation}
The FIRAS spectrum in Figure \ref{fig:residual} shows that 
$|y_D| < 1.5 \times 10^{-5}$ and $|\mu| < 9 \times 10^{-5}$
(95\% confidence),
and the spectral deviations corresponding to these limits are shown by
the solid and dashed curves.

The energy density transferred from the hotter electrons to the cooler
photons in the $y$ distortion is easily shown to be
$\Delta U/U = 4y < 6 \times 10^{-5}$.
For a $\mu$ distortion the corresponding limit is
\begin{equation}
\frac {\Delta U}{U} = 
\left(\frac {4\zeta(2)}{3\zeta(3)}-\frac{\zeta(3)}{\zeta(4)}\right) \mu
= 0.714\mu < 6 \times 10^{-5}.
\end{equation}
Thus any transfer of energy into the CMBR produced by scattering off of
hot electrons is less than 60 parts per million for all redshifts
less than $2 \times 10^6$, or times later than two months after the Big Bang.

\section{Anisotropy}

\subsection{History}

The earliest predictions of $\Delta T/T$ were very large.  
In addition to the $\Delta T/T = 3 \times 10^{-4}$ on $5^\prime$
scales predicted by Silk (1968),
Sachs and
Wolfe (1967) predicted $\Delta T/T = 1$\%  on larger angular scales
based on an assumed 
$\Delta \rho/\rho$ of 10\% over scales $L$ such that $H_\circ L = 0.1 c$.
The Sachs-Wolfe effect predicts $\Delta T/T = (1/3) \Delta \phi/c^2$,
where $\Delta \phi$ is the gravitational potential computed using
Newtonian gravity produced by the density fluctuations:
\begin{equation}
\nabla^2 (\Delta \phi) = 4\pi G \Delta \rho
\end{equation}

The first detection of anisotropy in the CMBR (Conklin 1969)
actually discovered a
different effect, the dipole anisotropy with a peak amplitude
of $\pm 0.12\%$ that is only indirectly related
to $\Delta \rho/\rho$.  
By 1971 a $3\sigma$ measurement (Henry 1971)
of the dipole anisotropy had been made,
and its significance was discussed by Peebles (1971).
The dipole anisotropy measures the velocity of
the observer relative to a very large piece of the Universe: a sphere
with comoving circumference $2 \pi (1+z_r) D_A(z_r)$,
where $D_A(z)$ is the angular size distance.
This velocity is produced by the action of the gravitational acceleration
$g = \vec{\nabla}(\Delta \phi)$ over the Hubble time $1/H_\circ$.
Further observations by Smoot, Gorenstein \& Muller (1977)
showed no detectable deviations from
the dipole pattern to a level below the expected $3 \times 10^{-4}$.

The analysis of anisotropy beyond the dipole is usually done in 
spherical harmonics, using
\begin{equation}
\frac{\Delta T(\hat{n})}{T_\circ} = \sum_{\ell,m} a_{\ell m} 
Y_{\ell m}(\hat{n})
\end{equation}
The rotational symmetry expected in the Universe means that all
the $a_{\ell m}$'s for a given $\ell$ should have the same variance,
and an expectation value of zero.  The variance of the $a_{\ell m}$'s
is $C_\ell$.  For a Harrison-Zel'dovich spectrum the angular power
spectrum is given by
\begin{equation}
C_\ell = \frac{4\pi \langle Q^2 \rangle}{5 T_\circ^2}\;\frac{6}{\ell(\ell+1)}
\end{equation}
The normalization is expressed in terms of $\langle Q^2 \rangle$, the
expected value of the quadrupole.  This often called $Q_{rms-ps}$, the
RMS quadrupole obtained by fitting to the power spectrum, or $Q_{flat}$.
The amplitude expressed as $\delta T_\ell$, which is the RMS $\Delta T$
produced by a all the harmonics in a band of width $\Delta \ell = \ell$,
is given by
\begin{equation}
\delta T_\ell = \sqrt{\frac{\ell(2\ell+1) T_\circ^2 C_\ell}{4\pi}}
\approx \sqrt{2.4}\; Q_{flat}
\end{equation}

Continued searches for anisotropy set upper limits
to the amplitude of a Harrison-Zel'dovich
spectrum of primordial density perturbations of
$\sqrt{\langle Q^2\rangle} < 55\;\mu$K  
in 1987 (Klypin {\it et al.} 1987)
and $\sqrt{\langle Q^2\rangle} < 22\;\mu$K 
in 1991 (Page, Cheng \& Meyer 1991).  The latter upper limit
is only 20\% higher than the eventual {\sl COBE\/} detection.

\subsection{DMR Observations}

The Differential Microwave Radiometers (DMR) experiment on {\sl COBE\/} was
designed to measure small temperature differences from place to place on
the sky.  The DMR consists of three separate units,
one for each of the three frequencies of
31.5, 53 and 90 GHz.  The field of view of each unit consists of two
beams that are separated by a $60^\circ$ angle that is bisected by the spin
axis.  Each beam has a $7^\circ$ FWHM.  The DMR is only sensitive to the
brightness difference between these two beams.  This differencing is
performed by a ferrite waveguide switch that connects the receiver input
to one horn and then the other at a rate of 100 Hz.  The
signal then goes through a mixer, an IF amplifier and a video detector.
The output of the video detector is demodulated by a lock-in amplifier
synchronized to the input switch.  The difference signal that results
is telemetered to the ground every 0.5 seconds.
Each radiometer has two channels: A and B.
In the case of the 31.5 GHz radiometer,
the two channels use a single pair of horns in opposite senses of
circular polarization.  In the 53 and 90 GHz radiometers there are 4 horns, 
and all observe the same sense of linear polarization.

A major contributor to the success of the DMR experiment was the four-fold
modulation a real cosmic signal had to display: the chop at 100 Hz, the
spin at 0.8 rpm, the orbital modulation at 0.01 rpm, and finally the annual
variation at 1 cycle per year.  Most successful anisotropy experiments have
at least three-fold modulation.  
For example, the Saskatoon experiment (Wollack {\it et al.} 1997)
chops with a big plate,
wobbles between East and West of the North celestial pole, and looks at the
daily modulation as the Earth turns.  The Tenerife experiment has four-fold
modulation since it chops rapidly between two horns, uses a plate to wobble,
and looks at both the daily and annual modulations by tracking a given
declination for a year.
Thus future missions should plan to modulate the true signals from the sky in
as many ways as are practical.

The major problems encountered by the DMR experiment were mainly Earth-related.
The Earth's magnetic field affected the ferrite Dicke switches, producing
a false signal which could be calibrated away.  The signal from the Earth's
limb, diffracted over the sunshade, was very difficult to determine as well as
a potentially damaging systematic error.
Thus future missions should look to observing sites well away from the Earth's
magnetic field and the Earth limb: somewhere in deep space, such as the
Earth-Sun L2-halo orbit, the Earth-Moon L4 or L5 point, or a heliocentric
orbit.

Wright (1996) extends the DMR sparse matrix  map-making technique 
to ``one-armed'' 
CMBR experiments with $1/f$ noise, using the time-ordered approach
developed by Wright, Hinshaw \& Bennett (1996) for differential radiometers.
Wright (1996) also gives more examples of systematic errors, and shows
how a complicated, multiply modulated scan pattern like the DMR's
4 way modulation can reduce the effect of systematic errors.

\subsection{DMR Results}

The basic DMR result is the discovery
of an intrinsic anisotropy (Smoot {\it et al.} 1992)
of the microwave background, beyond the
dipole anisotropy (Conklin 1969, Lineweaver {\it et al.} 1996).  
This anisotropy,
when a monopole and dipole fit to $|b| > 20^\circ$ are removed from
the map, and the map is then smoothed to a resolution of
$\approx 10^\circ$, is 30 $\mu$K.  The correlation function of this
anisotropy is well fit by the expected correlation function for
the Harrison-Zel'dovich
spectrum of primordial density perturbations
predicted by the inflationary scenario, and the amplitude 
($\sqrt{\langle Q^2 \rangle} = 17 \pm 5 \; \mu$K in
the first year data, and $18 \pm 1.6 \; \mu$K in the four
year data [Bennett {\it et al.} 1996]) is
consistent with many models of structure formation (Wright {\it et al.} 1992).

\subsection{Galactic Interference}

The emission from the galaxy shows a very strong dipole and quadrupole
pattern, so removing the galactic emission is essential for accurate
anisotropy measurements.  The separation of the observed signals into
galactic and cosmic components can be achieved using multifrequency data.
The spectrum of the cosmic signal is known: $\Delta I_\nu \propto
\partial B_\nu(T)/\partial T$, the spectrum of free-free emission
$\Delta I_\nu \propto \nu^{-2.1}$ is known, and the synchrotron and
dust components can be determined from maps at frequencies where
these components dominate the spectrum.

For the DMR data, with 3 frequencies, an internal linear combination
of the maps can be made that satisfies 3 criteria:
\begin{enumerate}
\item
A cosmic signal $\rightarrow \Delta T$,
\item
A free-free signal $\rightarrow 0$, and
\item
The observed spectrum of the galaxy $\rightarrow 0$.
\end{enumerate}
These three criteria uniquely specify the weights with which the
channels maps are combined:
$T_{NG} = -0.4512 T_{31} + 1.2737 T_{53} + 0.3125 T_{90}$.
Because of the high noise in the 31 GHz maps and the high coefficient
for the 53 GHz maps, this combination has a noise about twice as high as
the $53+90$ maps but much less response to galactic contamination.

The DMR data can be averaged into rings of
constant galactic latitude and fit to a linear function of $\csc|b|$.
The slope of $59\;\mu$K per unit $\csc|b|$ in the 31 GHz channels
is almost exactly what is expected if the
slope of $17\;\mu$K per unit $\csc|b|$ in the 53 GHz channels
were entirely due to free-free emission.  As a result this slope
is zeroed out in the no galaxy (NG) maps.
The slope {\it vs.}  $\csc|b|$ by itself contributes 10\% of the
observed variance of the 53 GHz sky smoothed to $10^\circ$ resolution.
This should be remembered as a cautionary note when using for
the ``COBE'' normalization a value of 
$\sqrt{\langle Q^2 \rangle}$ derived from
the $53+90$ GHz maps with no attempt to remove galactic emission.

For smaller angular scales the situation improves.
The dust emission has a power spectrum that follows $\ell^{-3}$
while the expected cosmic signal varies like $\ell^{-2}$.

\subsection{Power Spectrum}

\begin{figure}
\plotone{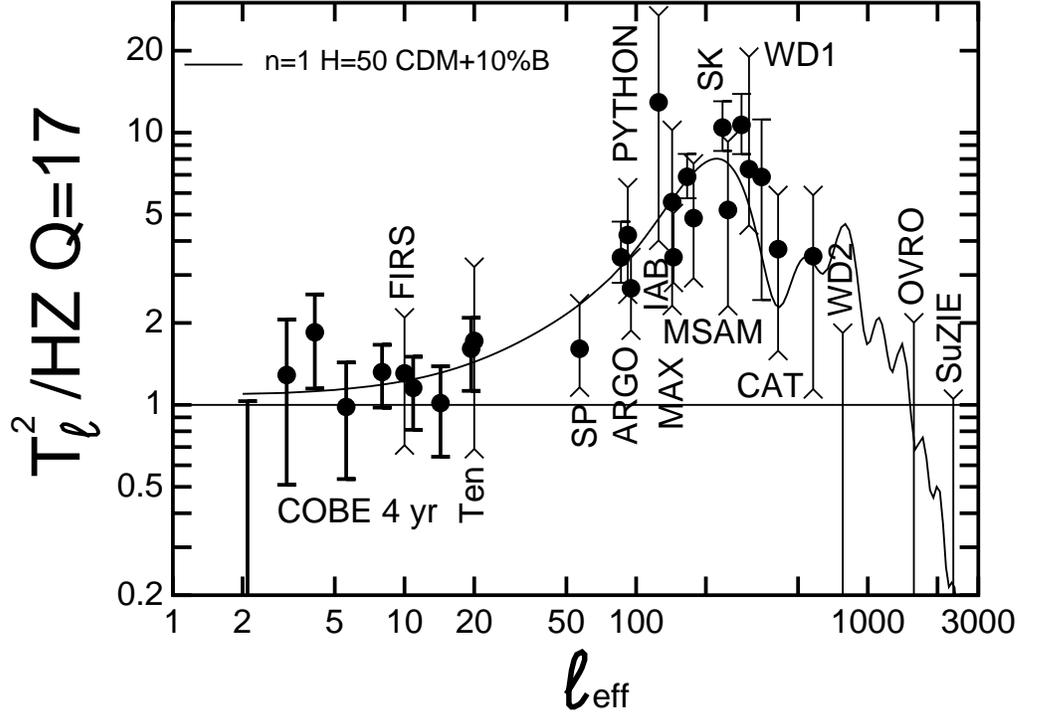}
\caption{The angular power spectrum of the CMBR from COBE and other
observations.  The COBE and Saskatoon points have straight ends,
while the other points have angled ends.  A CDM model with $n=1$ 
and $\Omega_B h^2 = 0.025$
is shown.\label{fig:lg4y5390}}
\end{figure}

The power spectrum of the DMR maps has been computed
by Wright {\it et al.} (1996) using
the Hauser-Peebles method to allow for incomplete sky coverage caused
by masking out the galactic plane region.
Figure \ref{fig:lg4y5390} shows the power in bands of $\ell$ normalized to
a pure $n = 1$ (Harrison-Zel'dovich) power law with 
the first year amplitude
$\sqrt{\langle Q^2 \rangle} = 17\;\mu$K, as computed by
Tegmark \& Hamilton (1997) using an improved quadratic estimator that
gives narrower window functions than the Hauser-Peebles method
and uncorrelated errors.
Also shown are results from various small and medium scale $\Delta T$
experiments:
FIRS (Ganga {\it et al.} 1994),
Tenerife (Hancock {\it et al.} 1997),
SP91 \& SP94 (Gundersen {\it et al.} 1995),
Saskatoon (Netterfield {\it et al.} 1997),
ARGO (Masi {\it et al.} 1996),
Python (Platt {\it et al.} 1997),
IAB (Piccirillo \& Calisse 1993),
MAX (Tanaka {\it et al.} 1996),
MSAM (Cheng {\it et al.} 1996),
CAT (Scott {\it et al.} 1996),
White Dish (Tucker {\it et al.} 1993),
OVRO (Readhead {\it et al.} 1989),
and SuZIE (Ganga {\it et al.} 1997).
For perturbations at scales smaller than the horizon at the end of radiation
dominance, dynamical effects boost the expected anisotropy.
At very small scales, the finite thickness of the recombination surface
filters out most of the anisotropy.
The model shown in Figure \ref{fig:lg4y5390}
is a CDM model computed by Sugiyama, taken from the Berkeley CMB server,
with $H_\circ = 50\;\mbox{km/s/Mpc}$, $\Omega_B h^2 = 0.025$, 
and the $n = 1$ expected from naive inflation.
The {\sl COBE\/} data is consistent with the apparent
spectral index $n_{app}$ slightly greater than the $n_{pri} \la 1$
that is predicted by inflation when the ``toe'' of the Doppler peak is
included.  
The 2-point correlation function of the DMR data (Hinshaw {\it el.} 1996a)
is consistent with the power spectrum.
Analysis with linear
(G\'orski {\it et al.} 1996, Hinshaw {\it et al.} 1996b) statistics instead
of the quadratic Hauser-Peebles statistics confirms this conclusion.

Note that the good agreement between the DMR
correlation function or power spectrum
and the Harrison-Zel'dovich model,
while supporting inflation, does not prove inflation, especially
since the precision with which the spectral index $n$ can be
determined is poor due to the small range of angular scales
probed by {\sl COBE.}  Thus intermediate scale anisotropy
data are needed for a more precise test of inflation.

The inflationary scenario also predicts that the temperature fluctuations
should have a Gaussian distribution.  This can be tested by looking at
higher-order moments of the maps such as three point correlation 
functions (Kogut {\it et al.} 1996a),
or by directly studying the probability distributions.  The random
measurement noise from the radiometers is still sufficient to interfere
with these studies to a great extent, but a preliminary analysis finds
that the Gaussian model is consistent with the observations.
The three-point correlation function of the 
{\sl COBE\/} DMR maps is
consistent with the level of three-point correlation expected in a
Harrison-Zel'dovich model with Gaussian fluctuations.
While the expected value of the three-point correlation vanishes for
Gaussian models, the variance does not, so any individual realization of
the Gaussian process will have a non-zero three-point correlation
function.

The CMBR field is moving so rapidly that any review article
is quickly out-of-date, but
White, Scott \& Silk (1994) is a review of the literature on
CMBR anisotropies and their interpretations up to early 1994.
Remember that all of the detections on Figure \ref{fig:lg4y5390} 
are less than five years old.

\section{Comparison to Large Scale Structure}

Wright {\it et al.} (1992) discussed the implications of the {\sl COBE\/} DMR
data for models of structure formation, and selected 4 models
from a large collection
given by Holtzman (1989) for detailed discussion:
a ``CDM'' model, with 
$H_\circ = 50\;\mbox{km/s/Mpc}$, $\Omega_{CDM} = 0.9$,
and $\Omega_B = 0.1$;
a mixed ``CDM+HDM'' model, with 
$H_\circ = 50\;\mbox{km/s/Mpc}$, $\Omega_{CDM} = 0.6$,
$\Omega_{HDM} = 0.3$ (a 7 eV neutrino), and $\Omega_B = 0.1$;
an open model, with 
$H_\circ = 100\;\mbox{km/s/Mpc}$, 
$\Omega_{CDM} = 0.18$,
and $\Omega_B = 0.02$; and
a vacuum dominated model, with 
$H_\circ = 100\;\mbox{km/s/Mpc}$,
$\Omega_{CDM} = 0.18$,
$\Omega_B = 0.02$, and $\Omega_{vac} = 0.8$.
The vacuum dominated model and especially the open model
have potential perturbations now that are too small to explain the
POTENT bulk flows (Bertschinger {\it et al.} 1990).

\begin{figure}
\plotone{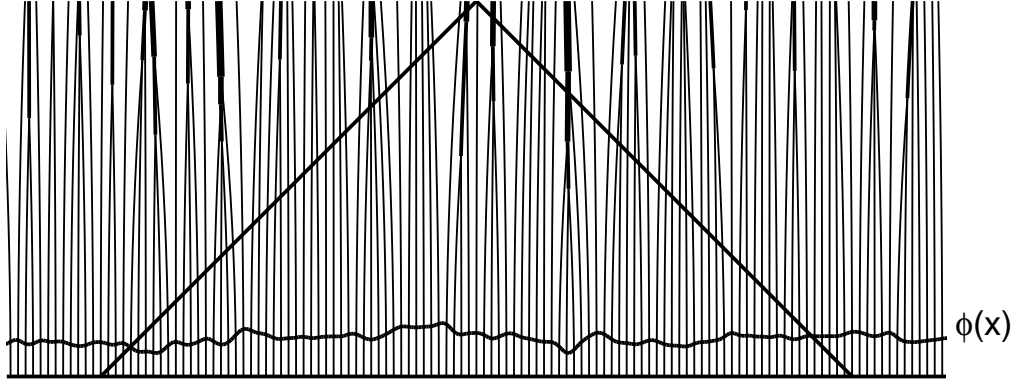}
\caption{The potential $\phi$ observed using the CMBR perturbs the
worldlines of galaxies in this conformal space-time diagram,
producing cluster of galaxies in the ``valleys'' and
voids in the ``mountains'' of the Universe.\label{fig:adhesion}}
\end{figure}

A comparison of the extremely large scale structure seen by {\sl COBE,}
to the large scale structure seen in studies of the clustering of
galaxies also leads to an estimate of the primordial spectral index,
$n_{pri}$.  The uncertainty
in this method is decreased because of the large range of scales
covered, but also increased due uncertainties in the models of
large scale structure formation.  However, this comparison strongly
favors $n = 1$.  Prior to the {\sl COBE,} announcement of anisotropy,
Peacock (1991) gave a implicit prediction that for $n = 1$ the
amplitude of $\Delta T$ should be 
$\sqrt{\langle Q^2 \rangle} = 18.8 \; \mu$K.
Peacock \& Dodds (1994) have extended this analysis of large scale
structure and I get a result $n_{pri} = 0.99 \pm 0.16$ from their paper
after correcting for their incorrect 
$\sqrt{\langle Q^2 \rangle} = 15 \; \mu$K and increasing the
uncertainty to allow for the uncertainty in the IRAS bias, $b_I$.
This result assumes that $\Omega = 1$, but Peacock \& Dodds have
also found that $\Omega^{0.6}/b_I = 1.0 \pm 0.2$.

So the basic question -- can gravity with a strength indicated by the
CMBR $\Delta T$ produce the the observed large scale structure? --
has the answer: Yes - but only if most of the matter in the Universe
is dark!  Figure \ref{fig:adhesion} shows how the CMBR
gives $\phi(x)$ which then drives the matter into clusters.

\section{Future CMB Work}

Jungman {\it et al.} (1996) show that many cosmological parameters
have an effect on the power spectrum of the CMB.  For example, the
value of $\Omega_\circ$ determines the location of the Doppler peak
at $\ell \approx 200$ in Figure \ref{fig:lg4y5390}, while the baryon density
affects its height.  
The position of the peak shifts to $\ell \approx 220/\sqrt{\Omega_\circ}$
in open models, and even today's non-systematic collection of competing
experiments shown in Figure \ref{fig:lg4y5390} suggests that open models
with $\Omega_\circ \approx 0.3$ can be ruled out.
The nature of the dark matter has an effect on the
secondary Doppler peaks, since hot dark matter will stream out of the
smaller scale structures whose ``bouncing'' causes the peaks.  Current
measurements of the small angular scale anisotropy of the CMB are
limited by their small sky coverage to relatively poor precision.  But
a new generation of proposed experiments can improve this situation 
dramatically.  With full sky coverage (actually only the usable 8~sr
away from the galactic plane) and a small instrument beam one can
achieve very high accuracy.
In bins with width $\Delta \ell/\ell =
0.2$, a 1\% uncertainty
requires mapping at least 50,000 beam areas, which is the whole sky 
for scales $\geq 0.9^\circ$.  Mapping the whole sky with this resolution 
will require a new satellite.  For smaller beams, long duration balloons and
observations from the South Pole will play a role along with satellites.
New space missions being planned include the ESA mission {\sl PLANCK},
formerly {\sl COBRAS/SAMBA},
and the US mission {\sl MAP} (Microwave Anisotropy Probe).
{\sl MAP\/} will have beam sizes in the $0.2 - 0.9^\circ$
range, which is 10 or more times smaller than the {\sl COBE\/} DMR beam.
{\sl PLANCK\/} will use bolometers to observe the CMBR at shorter wavelengths
and thus achieve beams sizes down to $0.07^\circ$, which is 100 times
smaller than the DMR beam.
If successful, these satellites will provide data with accuracy
better than 1\% out to $\ell > 600$ in the case of {\sl MAP\/},
and out to $\ell > 2000$ in the case of {\sl PLANCK\/}.
New balloon experiments being planned
include the US-Italian BOOMERANG mission and the US TOPHAT mission.
When these precise new anisotropy measurements are obtained, they
will establish several new constraints that cosmological models must
satisfy.

\section{Discussion}

The next decade of CMBR work should be a very active one.
The {\sl MAP\/} satellite is under construction and should be
launched in the year 2000.  
By 2002 data from {\sl MAP\/} should tell us:
\begin{enumerate}
\item Whether any of the CDM dominated models with adiabatic 
perturbations collapsing due to gravity actually agree with
the observations, and if so
\item Determine the values of the cosmological parameters $H_\circ$,
$\Omega_\circ$, $\Omega_B$, the cosmological constant, the neutrino masses, 
and the amplitude and spectral index $n$ of the primordial perturbations
produced in the first picosecond after the Big Bang.
\end{enumerate}
The {\sl PLANCK\/} mission will be launched a few years later, but with
its smaller beams and higher sensitivity it will be able to improve
the determinations of cosmological parameters from about 10\% accuracy
down to 1\% accuracy.
And thus in less than one century cosmology will change from a speculative
science with almost no observational data to an empirical science
very tightly constrained by precise observations.
These precise CMBR observations will reveal to us the nature of
the seeds of galaxies.

\acknowledgments

Work at UCLA on the Microwave Anisotropy Probe is supported by NASA grant
NAG53252 from the Goddard Space Flgiht Center.

\end{document}